\documentstyle[12pt]{article}
\newcommand{\be}[1]{\begin{equation} \label{(#1)}}
\newcommand{\ee}{\end{equation}}
\newcommand{\ba}[1]{\begin{eqnarray} \label{(#1)}}
\newcommand{\ea}{\end{eqnarray}}
\newcommand{\nn}{\nonumber}
\newcommand{\rf}[1]{(\ref{(#1)})}
\def \lsim {\mbox{${}^< \hspace*{-7pt} _\sim$}}

\def\p{\prime}
\def\pmb#1{\setbox0=\hbox{#1}%
  \kern-.015em\copy0\kern-\wd0
  \kern.03em\copy0\kern-\wd0
  \kern-.015em\raise.0233em\box0 }
\def\rp{$R_p \hspace{-1em}/\ \  $}
\def\rpm{R_p \hspace{-0.8em}/\;\:}
\def \sw {\sin\!\theta^{}_W }
\def \cw {\cos\!\theta^{}_W }
\def \tw {\tan\!\theta^{}_W }
\def \sw2 {\sin^2\!\theta^{}_W }
\def \sbt {\sin\!\beta }
\def \s2bt {\sin\!2\beta }
\def \cbt {\cos\!\beta }
\def \lg  {\langle}
\def \rg  {\rangle}
\def \tbt {\tan\!\beta }

\def \znbb {0\nu\beta\beta}
\def\bfr{\pmb{${r}$}}
\def\bfsgm{\pmb{${\sigma}$}}
\def\sir{({ \bfsgm_{a}^{~}} \cdot {\hat{\bfr}_{ab}^{~}} )}
\def\sjr{({ \bfsgm_{b}^{~}} \cdot {\hat{\bfr}_{ab}^{~}} )}
\def\si{{ \bfsgm_{a}^{~}}}
\def\sj{{ \bfsgm_{b}^{~}} }
\begin{document}

\begin{center}
{\bf Bilinear R-parity Violation in Neutrinoless Double Beta Decay.}\\
\bigskip {Amand Faessler, Sergey Kovalenko\footnote{
On leave of absence from the Joint Institute for Nuclear Research, Dubna,
Russia.}\\
\bigskip {\it Institute f\"ur Theoretische Physik  der Universit\"at
T\"ubingen,\\[0pt]
Auf der Morgenstelle 14, D-72076 T\"ubingen, Germany}\\[0pt]
\bigskip Fedor \v Simkovic \\[0pt]
\bigskip {\it Department of Nuclear Physics, Comenius University, \\[0pt]
Mlynsk\'a dolina F1, 84215 Bratislava, Slovakia} }
\end{center}
\bigskip

\begin{abstract}
We discuss some phenomenological issues of the effective quark-lepton
operators emerging from the bilinear lepton-Higgs couplings in the
superpotential and in the soft supersymmetry (SUSY) breaking
sector of the supersymmetric models without R-parity.
The contribution of these operators
to the neutrinoless double beta decay ($\znbb $) is derived.

The corresponding nuclear matrix elements are calculated
within the renormalized quasiparticle random phase approximation,
which includes the Pauli effect of fermion pairs and does not collapse
for the physical values of the nuclear force strength.

On this basis we extract from the experimental data new stringent
limits on the 1st generation mass parameter characterized
the lepton-Higgs bilinear coupling and on the electron sneutrino
vacuum expectation value.
\end{abstract}

\section{Introduction}
In the standard model (SM) of the electro-weak interactions
the baryon B and lepton L numbers conservation is protected
to all orders of perturbation theory by an accidental
$U_{1B}\times U_{1L}$ symmetry existing at the level of
renormalizable operators.
In the minimal supersymmetric (SUSY) extension of the standard model
(MSSM) \cite{mssm} this symmetry is absent
and the L and B violating processes are not forbidden.
A conventional way of eliminating the phenomenologically dangerous
L,B-violation in this case exploits a discrete symmetry
known as $R$-parity \cite{Rp}, ~\cite{hall} which is imposed on
the model.
 This is a multiplicative $Z_2$
symmetry defined as
$R_p=(-1)^{3B+L+2S}$, where $S,\ B$ and $L$ are the spin,
the baryon and the lepton quantum numbers.
R-parity conservation has a distinctive
phenomenology. It prevents lepton and baryon number violating processes,
the superpartners are produced in associated production and
the lightest SUSY particle is stable. The latter leads to the celebrated
missing $E_T$ signature of the SUSY event in high energy detector
and renders a cold dark matter particle candidate.
Although desirable for many reasons the R-parity conservation has no well
motivated theoretical grounds.

On the other hand relaxing the R-parity conservation we may get
a new insight into the long standing problems of particle physics,
in particular, to the neutrino mass problem. Remarkable,
that in this framework neutrino can acquire the tree level
supersymmetric mass via the mixing with the gauginos and
higgsinos at the weak-scale \cite{hall}, \cite{nilles}-\cite{bgnn96}.
This mechanism does not involve the physics at
the large energy scales $M_{int} \sim {\cal O}(10^{12}$GeV)
in contrast to the see-saw mechanism
but relates the neutrino mass to the weak-scale physics accessible
for the experimental searches.

The R-parity can be broken (\rp) either explicitly \cite{hall}
or spontaneously \cite{spont}. The first option allows one
to establish the most general phenomenological consequences of
R-parity violation while a predictive power in this case is
rather weak due to the large number of free parameters.
Spontaneous realization of \rp SUSY is much more predictive
scheme leading to many interesting phenomenological consequences
\cite{Valle-ring}.
However, it represents a particular model of the R-parity violation.
At present it is an open question which underlying high-energy
scale physics stands behind the R-parity, protecting or violating it
at the weak scale.

Many aspects of the \rp SUSY models in high and low energy
processes had been investigated in the
literature \cite{hall}-\cite{Rp-status}, \cite{Mohapatra}-\cite{FKSS97}.

Recently, a growing interest to the supersymmetric models
without R-parity was stimulated by the exciting news from the
HERA experiments, reported the anomaly in deep inelastic
$e^+p$-scattering \cite{hera} which can be elegantly explained within
these theoretical framework in terms of
the lepton number violating interactions.

Since the lepton number is not conserved without R-parity some low-energy
exotic processes become possible within the \rp MSSM.
Among them the neutrinoless nuclear double beta decay ($\znbb$)
is known to be very sensitive to the certain \rp interactions
\cite{HKK96}. Provided an unprecedented accuracy of the modern
$\znbb$-decay experiments \cite{hdmo97} this allows one to establish
stringent constraints on the \rp SUSY \cite{Mohapatra}-\cite{FKSS97}.

In the present paper we consider the implications of the bilinear
lepton-Higgs \rp terms on $\znbb$-decay. In the general  case of
the explicitly broken R-parity these terms are present in
the superpotential and in the soft SUSY breaking potential.
Previously the main attention was paid to the  phenomenology of
the trilinear \rp Yukawa couplings. It was widely believed that
the bilinear
\rp terms can be rotated away by a proper field redefinition.
However, it is not the case in the presence of the soft SUSY breaking
interactions \cite{cgjrj95}, \cite{Valle-ring}.
It was realized that the bilinear \rp violation,
generically leading to the non-zero vacuum expectation values (VEV) of
the sneutrino fields and to the lepton-gaugino-higgsino and
slepton-Higgs mixing, provides a number of interesting
phenomenological issues
\cite{nilles}-\cite{bgnn96}, \cite{LH-ph}-\cite{Mukh}.

In particular, this mixing generates the new effective lepton number
violation operators which
contribute to the nuclear $\znbb$-decay. In what follows we derive these
operators and analyze their net effect in the presence of the nuclear
media.

The paper is organized as follows.
Basic ingredients of the \rp MSSM with the general setting of
the explicit R-parity violation are shortly described in Section 2.
In Section 3 we discuss the bilinear \rp mechanism of the nuclear
$\znbb$-decay. Here we analyze all the tree-level \rp MSSM contributions
to the $\znbb$-decay amplitude.
We start with the quark level and derive the corresponding low energy
effective Lagrangian. In Section 4 we take into account
the effect of nuclear structure and derive the corresponding
nuclear matrix elements. Then we calculate their values within
the renormalized Quasiparticle Random Phase Approximation
(pn-RQRPA) \cite{pn-RQRPA}.
The pn-RQRPA is an extension of the pn-QRPA by taking into account
the effects of the Pauli principle for the fermion pairs.
In this approach the sensitivity of the nuclear matrix elements to
the details of the nuclear Hamiltonian is reduced considerably.
Using experimental lower bound on the $^{76}$Ge half-life we extract
in Section 5 stringent constraints on the 1st generation
lepton-Higgs mixing mass parameter and on the electron sneutrino VEV.
We close our discussion with the short comments on some implications of
these constraints for the other experiments.

\section{Minimal SUSY model with R-parity violation.}
In order to set up our notations let us briefly recapitulate
the main ingredients of the  minimal SUSY standard model (MSSM)
with explicit R-parity violation (\rp MSSM).

The $R_p$ violation is introduced into the theory through
the superpotential and soft SUSY breaking sector.

For the minimal MSSM field contents the most general
gauge invariant form of the renormalizable superpotential reads
\ba{sup_gen}
           W = W_{R_p} + W_{\rpm}.
\ea
The $R_p$ conserving part has the standard MSSM form
\be{R_p-cons} %1
         W_{R_p} = h_L H_1 L E^c + h_D H_1 Q D^c
                   + h_U H_2 Q U^c + \mu H_1 H_2.
\ee
Here $L$,  $Q$  stand for lepton and quark
doublet left-handed superfields while $ E^c, \  U^c,\   D^c$
for lepton and {\em up}, {\em down} quark singlet  superfields;
$H_1$ and $H_2$ are the Higgs doublet superfields
with a weak hypercharge $Y=-1, \ +1$, respectively.
Summation over the generations is implied.

The $R_p$ violating part of the superpotential \rf{sup_gen}
can be written as ~\cite{Rp}, ~\cite{hall}
\ba{W_rp}
W_{\rpm} = \lambda_{ijk}L_i L_j E^c_k +
\lambda^{\prime}_{ijk}L_i Q_j D^c_k + \mu_i L_j H_2+
\lambda^{\prime\prime}_{ijk} U^c_i D^c_j D^c_k,
\ea
The coupling constants $\lambda$ ($\lambda^{\prime\prime}$) are
antisymmetric in the first (last) two indices.
The first two terms violate
lepton number while the last one violates baryon number conservation.

Another source of the R-parity violation is the soft supersymmetry
breaking part of the scalar potential. It contains the \rp-terms
\ba{V_rp}
V_{\rpm}^{soft} = \tilde\lambda_{ijk}\tilde L_i \tilde L_j
\tilde E_k^c +
\tilde\lambda^{\prime}_{ijk}\tilde L_i \tilde Q_j \tilde D_k^c +
\tilde\lambda^{\prime\prime}_{ijk}\tilde U_i^c \tilde U_j^c \tilde D_k^c +
\tilde\mu_i\tilde L_i H_2 +&& \\ \nn
+ m^2_{LH}\tilde L_i H^{\dagger}_1 + \mbox{H.c.}&&
\ea
The simultaneous presence of lepton and baryon number violating
terms in Eqs. \rf{W_rp}, \rf{V_rp} (unless the couplings are very small)
would cause unsuppressed proton decay. Therefore, either the lepton or
the baryon number violating couplings can be present. There may exist
in the theory an underlying discrete symmetry such as the B-parity
\cite{hall}, \cite{B-parity} which forbids dangerous combinations of
these couplings.
Henceforth we simply set
$\lambda^{\prime\prime}= \tilde\lambda^{\prime\prime}=0$.

The remaining R-parity conserving part of the soft SUSY breaking
sector includes the scalar field interactions
\ba{V_Soft}
V_{R_p}^{soft}= \sum_{i=scalars}^{}  m_{i}^{2} |\phi_i|^2 +
h_L A_L H_1 \tilde L \tilde{E}^c + h_D A_D H_1 \tilde Q \tilde{D}^c -&&
\\  \nn
- h_U A_U H_2 \tilde Q \tilde{U}^c  -
\mu B H_1 H_2 + \mbox{ H.c.}&&
\ea
and the "soft"  gaugino mass terms
\be{M_soft}
{\cal L}_{GM}\  = \ - \frac{1}{2}\left[M_{1}^{} \tilde B \tilde B +
 M_{2}^{} \tilde W^k \tilde W^k  + M_{3}^{} \tilde g^a \tilde g^a\right]
 -   \mbox{ H.c.}
\ee
As usual, $M_{3,2,1}$ denote the masses of the $SU(3)\times
SU(2)\times U(1)$ gauginos $\tilde g, \tilde W, \tilde B$ while $m_i$
stand for the masses of the scalar fields.
The gluino $\tilde g$ soft mass $M_3$
coincides in this framework with its physical mass denoted
hereafter as $m_{\tilde g} = M_3$.
$A_L,\ A_D, \ A_U$ and $B$ in Eq. \rf{V_Soft} are
trilinear and bilinear "soft" supersymmetry breaking parameters.
All these quantities are free SUSY model parameters which due to
the renormalization effect depend on the energy scale.

In this paper we assume for simplicity the universal gaugino soft
masses at the grand unification scale $M_{GUT}$.
At the weak scale this leads to the following relations
\ba{gut}
M_1 = (5/3) \tan\theta_W^2 M_2, \ \ \ M_2 \approx 0.3 M_3,
\ea

An impact of the R-parity violation on the low energy phenomenology is
twofold. First, it leads the lepton number (LNV) and lepton flavor (LFV)
violating interactions directly from the trilinear terms in
$W_{\rpm}$. Second, bilinear terms in $W_{\rpm}$ and
in $V_{\rpm}^{soft}$ generate the non-zero vacuum
expectation value for the sneutrino fields
$\langle\tilde\nu_i\rangle\neq 0$ and cause
neutrino-neutralino as well as electron-chargino mixing. The mixing brings
in the new LNV and LFV interactions in the physical mass eigenstate basis.
Below we will specify those interactions which are relevant for
the $\znbb$-decay.

The trilinear terms of the R-parity breaking part of the superpotential
$W_{\rpm}$ lead to the following $\Delta L=1$ lepton-quark operators
\ba{lambda}
{\cal L}_{\lambda } &=&
\lambda _{ijk}[\tilde{\nu}_{iL} \bar{e}_{k} P_L e_{j}+
\tilde{e}_{jL}\bar{e}_{k} P_L \nu _{i}+
\tilde{e}_{kR} \bar{e}_{j} P_R \nu^c_i -(i\leftrightarrow j)]\ + \\ \nn
&+&\lambda _{ijk}^{\prime }[\tilde{\nu}_{iL}\bar{d}_{k} P_L d_{i}+
\tilde{d}_{jL}\bar{d}_{k} P_L \nu _{i} +
\tilde{d}_{kR}\bar{d}_{j} P_R \nu^c_{i}-
\tilde{e}_{iL}\bar{d}_{k} P_L u_{j}   \\ \nn
&-&\tilde{u}_{jL}\bar{d}_{k} P_L e_{i}-
\tilde{d}_{kR} \bar{u}_{j} P_R e^c_i]\ +\ \mbox{H.c.}
\ea
Here, as usual $P_{L,R} = (1 \mp \gamma_5)/2$.

The presence of the bilinear terms in the Eqs. \rf{W_rp},\rf{V_rp}
leads to the terms in the scalar potential linear in the sneutrino
fields $\tilde{\nu}_{i}$. As a result, at the minimum of the potential
$\langle\tilde{\nu}_{i}\rangle\neq 0$. Thus, the MSSM vertices
$\tilde{Z}\nu \tilde{\nu}$ and  $\tilde{W}e\tilde{\nu}$
create the gaugino-lepton mixing mass terms
$\tilde{Z}\nu\langle\tilde{\nu}\rangle,
\tilde{W}e\langle\tilde{\nu}\rangle$
(with $\tilde{W},\tilde{Z}$ being wino
and zino fields). Combining this terms with the lepton-higgsino
$\mu_i L_i\tilde{H}_{1}$ mixing from the superpotential Eq. \rf{W_rp}
we end up with
%come up with
$7\times 7$ neutral fermion and $5\times 5$ charged fermion mass matrices
\mbox{(see Appendix A).}
The mass eigenstate fields can be written in the form
\ba{mass}
\Psi_{(0)i} = \Xi_{ij}\  \Psi_{(0)j}^{\p},\ \ \ \ \ \ \
\Psi_{(\pm)i} = \Delta_{ij}^{\pm}\  \Psi_{(\pm)j}^{\p},
\ea
with the weak eigenstate fields in two component notation
\ba{weak}
\Psi_{(0)}^{\p T} &=& (\nu_i,\,\,-i\lambda^{\p },\,\,
-i\lambda _{3},\,\, \tilde{H}^0_1,\,\, \tilde{H}_2^0), \\
\Psi_{(-)}^{\p T} &=& (e^-_L,\,\, \mu^-_L,\,\, \tau^-_L,\,\,
-i\lambda_{-},\,\, \tilde H^-_1),\\
\Psi_{(+)}^{\p T} &=& (e^+_L,\,\, \mu^+_L,\,\, \tau^+_L,\,\,
-i\lambda_{+},\,\, \tilde H^+_2).
\ea
Here $\nu_i$ are the neutrino fields, $\lambda^{\p}$ and
$\lambda_{3}, \lambda_-$ are the $U_{1Y}$ and $SU_{2L}$
gauginos, respectively while higgsinos are denoted as
$\tilde H^0_{1,2}, \tilde H^{\pm}_{1,2}$.  The mixing matrices
$\Xi$ and $\Delta^{\pm}$ diagonalize the neutralino-neutrino and
the chargino-charged lepton mass matrices respectively.
The lightest mass eigenstates are identified with the physical neutrinos
and the charged leptons. Remarkable, that as a result of the minimal
field content and the gauge invariance the neutral fermion mass matrix
${\cal M}_0$ \rf{neut_m} before diagonalization has such a texture that
its first three rows and the last one are
linearly dependent and, as a result, two neutrino mass eigenstates
are degenerate massless states. The third neutrino state acquires
the tree level mass which approximate form is (see Appendix B)
\ba{nu_m}
m_{\nu} = \frac{2}{3} \frac{g^{2}_1 M_2}{Det M_{\chi}}
|\vec{\Lambda}|^2,
\ea

It is natural to identify the massive neutrino state with the tau neutrino
$\nu_{\tau}$ while the two massless states with the $\nu_e$ and
$\nu_{\mu}$. The $\nu_{e}-\nu_{\mu}$ mass degeneracy is lifted by
the 1-loop corrections as well as by the non-renormalizable terms in
the superpotential giving to $\nu_{e,\mu}$ the small non-equal masses
\cite{bgnn96}.
As to the tau neutrino mass in Eq. \rf{nu_m} it is subject to
the experimental constraint $m_{\nu_{\tau}} \leq 23$MeV \cite{nu_exp}.
Assuming no cancellation in Eq. \rf{nu_m} this leads to the
upper bounds
\ba{tau_const}
\mu_i \ \ \ \lsim \ \ \  15\mbox{GeV}, \ \ \ \ \ \
\langle\tilde\nu_i\rangle\ \ \  \lsim \ \ \ 7\mbox{GeV}.
\ea
at the typical sample values of the MSSM parameters
$\mu \sim M_2 \sim M_W$. Of course, these bounds are only indicative
and may essentially vary from point to point in the MSSM parameter space.

The $m_{\nu_{\tau}}$ constraints can be evaded assuming an approximate
alignment between two vectors
$a_i = (\mu_i, \mu)$ and $b_i = (\lg\tilde\nu_i\rg, \lg H_1^0\rg)$
which leads to the cancellation in Eq. \rf{nu_m} since
$|\vec{\Lambda}|^2 = |\vec a|^2|\vec b|^2 -
\left(\vec a\cdot \vec b\right)^2$. This might be guaranteed by a special
global symmetry \cite{banks} or by some dynamical reasons \cite{nilles}.

Rotating the MSSM Lagrangian to the mass eigenstate basis one obtains
the RPM generated lepton number violating interactions which bring many
interesting implications for the low and high energy phenomenology.
Below we are studying they contribution to the $\znbb$-decay.

\section{LH-induced $\znbb$-decay. Quark level transitions.}
We have analyzed all the possible tree level contributions
to the $\znbb$-decay amplitude which include the RPM interactions and
the superpotential $\lambda, \lambda'$ couplings
from Eq. \rf{lambda}. The leading diagrams are presented in the Fig. 1.
The diagrams in Fig.1(a,b) incorporate only the RPM generated
vertices, and in Fig. 1(c,d) these vertices are accompanied by one
$\lambda^{\p}$ type vertex (on the top of the diagrams).
The diagram in Fig. 1(a) has in
the intermediate state either neutrinos or neutralinos and two
W-bosons while the diagrams in Fig. 1(c,d) neutrinos,
squarks/selectron and one W-boson. The diagram
in Fig.1(b) is mediated by the gluino and double squark exchange.
The diagram Fig. 1(a) with the neutrino exchange is the conventional
Majorana neutrino contribution to the $\znbb$-decay.
 Recall that in the \rp MSSM with the bilinear R-parity violation
the neutrino masses and mixing angles are derived at the tree level
in terms of $\mu_i, \lg\tilde\nu_i\rg$ and the MSSM parameters
(see Appendix A).
Therefore, this contribution inherently pertains to this model.
We did not include in this list those diagrams which do not contain
RPM vertices. These diagrams constructed of the $\lambda,\lambda^{\p}$
couplings were previously analyzed in
Refs. \cite{Mohapatra}-\cite{FKSS97}.
All the other diagrams in this order of perturbation theory have
extra suppression factors and, therefore, can be neglected.
The suppression factors originate from the smallness of neutrino mass,
when it appears in a positive power, from
the 1st generation left-right sfermion
mixing proportional to $m_{u,d,e}/M_{SUSY}$ and/or
from the fermion-sfermion-higgsino
couplings proportional to $m_{u,d,e}/M_W$ with $m_{u,d,e}$ being
the u,d quark and the electron masses respectively while $M_{SUSY}$
denotes the typical SUSY breaking mass scale.

Now let us specify those RPM generated operators
which are encountered in the diagrams in Fig. 1.  They are
\ba{LH-vert}
&&{\cal L}_{LH} = -\frac{g_2}{\sqrt{2}}
\kappa_n W^-_{\mu} \bar e \gamma^{\mu}P_L \chi_n +\\ \nn
&&+ \sqrt{2} g_2 \left(\beta_k^d\bar\nu_{_k} P_R d\ \tilde d_{_R}^* +
                           \beta_k^u\bar\nu_k P_R u^c\ \tilde u_{_L} +
         \beta_{ki}^e\bar\nu_k P_R e^c\ \tilde e_{_{Li}} +
\zeta \bar u P_R e^c\ \tilde d_{_L}\right) + H.c.
\ea
The subscripts $k,i$ denote generations.

The first term is generated from the standard model $W-e-\nu$ and
the MSSM $W-\chi^{\pm}-\chi$ interaction terms while the rest
originates from the MSSM neutralino (chargino)-fermion-sfermion
interactions $\chi-q-\tilde q$, $\chi^{\pm}-q-\tilde q$
(for the MSSM Lagrangian see \cite{mssm}).

Note that the trilinear fermion-sfermion couplings in ${\cal L}_{LH}$
are not present among the superpotential trilinear
$\lambda, \lambda^{\p}$ terms in Eq. \rf{lambda}.

The coefficients in Eqs. \rf{LH-vert} depend on the
mixing matrix elements introduced in Eq. \rf{mass}:
\ba{coeff_0}
\kappa_n &=& \sum_{k=1}^{3} \Delta^-_{11} \Xi_{n+3,3} +
\sqrt{2}\Delta^-_{14} \Xi_{n+3,5}  + \Delta^-_{15} \Xi_{n+3,6},\\ \nn
\beta_{ki}^e &=& -\frac{1}{\sqrt{2}}\Xi_{14} V^{(\nu)*}_{ik} +
\frac{1}{2}\left(\tw \Xi_{k4} + \Xi_{k5}\right)\delta_{i1},\\ \nn
\beta_k^u &=& -\frac{1}{6}\left(\tw \Xi_{k4} + 3 \Xi_{k5}\right),\\ \nn
\beta_k^d &=& -\frac{1}{3}\tw \Xi_{k4},\ \
\zeta = -\frac{1}{\sqrt{2}} \Delta^-_{14}.
\ea
In what follows,  for the derivation of the constraints on
parameters $\langle\nu_i\rangle, \mu_i$ characterizing
the bilinear \rp we employ the approximate analytical
diagonalization method of the Ref. \cite{Now}. It allows one
to represent the mixing
matrices in a convenient analytic form and express the dependence of
the coefficients in Eqs. \rf{coeff_0} on the afore-mentioned \rp parameters
explicitly.
In the leading order in small parameters
$\mu_i/M_Z$, \mbox{$\langle\tilde\nu_i\rangle/M_Z$}
we obtain
\ba{coeff}
\kappa_n &=& \xi^*_{1k} N^*_{nk} - \sqrt{2}\xi^L_{11} N^*_{n2}
-\xi^L_{12} N^*_{n3},\\ \nn
\beta_{ki}^e &=& \frac{1}{\sqrt{2}}\xi^L_{11} V^{(\nu)*}_{ik} -
\frac{1}{2}V^{(\nu)*}_{jk}\left(\tw \xi^*_{j1} +
                               \xi^*_{j2}\right)\delta_{i1},\\ \nn
\beta_k^u &=& \frac{1}{6} V^{(\nu)*}_{jk} \left(\tw \xi^*_{j1} +
                                            3 \xi^*_{j2}\right),\\ \nn
\beta_k^d &=& \frac{1}{3} \tw V^{(\nu)*}_{jk} \xi^*_{j1},\ \
\zeta =  \frac{1}{\sqrt{2}} \xi^L_{11}.
\ea
The notations used in these formulas are explained in Appendix B.

The MSSM gluino-quark-squark vertex in the diagram Fig. 1(b)
is described by the Lagrangian term
\ba{gluino}
                {\cal L}_{\tilde g} = - \sqrt{2} g_3
        \frac{{\bf \lambda}^{(a)}_{\alpha \beta}}{2}
        \left( \bar q^{\alpha} P_R \tilde g^{(a)} \tilde q_L^{\beta}
                 - \bar q^{\alpha} P_L \tilde g^{(a)} \tilde q_R^{\beta}
                        \right)
			+ h.c.,
\ea
Here ${\bf \lambda}^{(a)}$ are $3\times 3$ Gell-Mann matrices
($a = 1,..., 8$). Superscripts $\alpha, \beta$ denote the color indices.

The diagrams in Fig.1 describe the \rp SUSY induced quark
transitions which proceed in the nuclear media and trigger
the nuclear $\znbb$-decay. Our goal is to derive
the corresponding half-life for a certain isotope
assuming for simplicity that there is no other
contributions to this nuclear process.
In order to apply the standard approach \cite{HKK96},
\cite{FKSS97}, \cite{doi85}  based on
the non-relativistic impulse approximation one has to
derive first the effective low energy Lagrangian describing the
basic $\znbb$-quark transition $d d\longrightarrow u u e e$
in terms of the color singlet quark charged currents
which can be embedded then into the corresponding hadronic
(nucleon or pi-meson) currents inside a nucleus. One has also
to separate the short and long ranged parts of the quark level
transition operators since they are treated within this approach
in different ways. It is understood that the short ranged parts
involve only heavy  particles in the intermediate states
($\chi, \chi^{\pm}, W, \tilde q, \tilde e$)
while the long distance ones include the neutrino exchange.

Integrating out the heavy fields from the diagrams in Fig. 1
and carrying out Fierz reshuffling we obtain the desired effective
Lagrangian which allows one to reproduce the low energy contribution of
these diagrams in the first or in the second order of
perturbation theory. It takes the form
\ba{eff}
{\cal L}_{eff}(x)\ &=&\
\frac{G_F^2}{2 m_P}
\left[\eta_{\tilde g}(J\  J  -
\frac{1}{4} J^{\mu\nu}\ J_{\mu\nu})
 + \eta_{\chi} J^{\mu}J_{\mu}\right]
(\bar e P_R e^{\bf c}) - \\  \nn
&-& \sqrt{2}G_F \lambda_{i11}^{\p}\cdot
\eta^{(ki)}_{\lambda}\cdot \left(\bar \nu_k P_R e^c\right) J +
G_F \sqrt{2}(\bar e \gamma^{\mu} P_L \nu_k)V^{(\nu)}_{1k} J_{\mu}.
\ea
Here we introduced the color singlet quark currents
\ba{currrents}
J = \bar u^{\alpha} P_R d_{\alpha},\ \ \
J^{\mu \nu} = \bar u^{\alpha} \sigma^{\mu \nu} P_R d_{\alpha},\ \ \
J^{\mu} = \bar u^{\alpha} \gamma^{\mu}P_L d_{\alpha},
%\label{(JT)}
\ea

The effective parameters $\eta$ accumulating the dependence on
the initial \rp SUSY parameters are defined as
\ba{eta}
\eta_{\tilde g} &=& \frac{4\pi\alpha_s}{9}
\frac{g_2^2\zeta^2}{G_F^2 m_{\tilde d_L}^4}
\left(\frac{m_{_p}}{m_{\tilde g}}\right); \ \
\eta_{\chi} = \sum_{i=1}^4 \frac{m_{_p}}{m_{\chi_i}} \kappa_i^2 \equiv
\frac{m_{_p}}{\langle m_{\chi}\rangle},\\ \nn
\eta^{(ki)}_{\lambda} &=&
\frac{g_2}{2 G_F} \left(2 \frac{\beta_{ki}^e}{m_{\tilde e_{Li}}} -
\frac{\beta_k^d}{m_{\tilde d_R}}\delta_{i1} -
\frac{\beta_k^u}{m_{\tilde u_L}}\delta_{i1}\right).
\ea
Here $m_{\tilde g}, m_{\tilde q}$ and $m_{\chi_i}$ are the gluino,
squark and neutralino masses.

In the \rp MSSM we have for the neutrino mixing
matrix element (see Appendix B) the following expression
\ba{nu-mix}
V^{(\nu)}_{1k} =
\delta_{1k}\cos\theta - \delta_{3k}\sin\theta
\ea
with
\ba{theta}
\sin\theta = - \Lambda_1/|\vec\Lambda|.
\ea

In Eq. \rf{eff} the first and the second terms reproduce the contribution
of the gluino Fig. 1 (b) and the neutralino Fig. 1(a) exchange graphs
in the 1st order of perturbation theory while the third and the last terms
reproduce the contribution of the neutrino exchange graphs in
Fig. 1(a) and Fig. 1(c,d) in the 2nd order of perturbation theory.
Note, that the last term in Eq.\rf{eff} is the ordinary lepton-number $L$
conserving standard model interaction term. Since the $\znbb$-decay
requires the L-violation $\Delta L = 2$ the neutrinos
in Fig. 1 (a) propagate in the Majorana lepton-number violating mode.
In this case the source of the L-violation is given by
the neutrino Majorana mass term. That is why the contribution
corresponding to this diagram is proportional to the neutrino mass
or, more precisely, to the average neutrino mass
$\lg m_{\nu}\rg$ defined below. On the contrary, the neutrino exchange
diagrams in Fig. 1 (c,d) are not proportional to $\lg m_{\nu}\rg$ and
survive in the limit $m_{\nu} = 0$ since the lepton-number
violation $\Delta L = 2$ is produced by the interaction term
$\bar\nu P_R e^c$ in Eq. \rf{eff} itself. Therefore, the neutrinos
in diagrams Fig. 1 (c,d) propagate in the L-conserving Dirac mode.

So far we concentrated on the $\znbb$-transitions
at the quark level described by the effective Lagrangian
\rf{eff}.
The aim of this paper is the calculation of
the amplitude for the nuclear $\znbb$-decay taking into account
nuclear structure.

The next section deals with the derivation of the amplitude for
the nuclear $\znbb$-decay triggered by the quark transitions in Fig.1.

\section{Nuclear $\znbb$-decay. }

Let us write down the following formal expression for the amplitude
$R_{0\nu\beta\beta}$  of $0\nu\beta\beta$-decay
\ba{DefL}
{\cal R}_{\znbb} \ &=& <(A, Z + 2), 2 e^-| S - 1|(A, Z)> = \\ \nn
&=& <(A, Z + 2), 2 e^-| T exp[i \int d^4 x {\cal L}_{eff}(x)] |(A, Z)>
\ea
where the effective Lagrangian ${\cal L}_{eff}$
is given by Eq. \rf{eff}.
The nuclear structure is involved via the initial (A,Z) and
the final (A, Z+2) nuclear states having the same atomic
weight A, but different electric charges Z and Z+2.
The standard framework for the calculation of this nuclear matrix
element is the non-relativistic impulse approximation (NRIA) \cite{doi85}.

It is straightforward to derive the following formula
for the amplitude of the  $0^+\rightarrow 0^+$ transition
with two outgoing S-wave electrons
\ba{R_tot}
&&{\cal R}_{\znbb}(0^+ \rightarrow 0^+) =
C_{0\nu} f_A^2\bar e (1 + \gamma_5) e^c\times \\ \nn
&&\left[\eta_{\tilde g}{\cal M}_{\tilde g} + \lambda_{111}^{\p}\eta^{(k1)}_{\lambda} V^{(\nu)}_{1k} {\cal M}_{\lambda} +
 \frac{m_p}{<m_{\chi}>} {\cal M}_{N} +
\frac{<m_{\nu}>}{m_e}{\cal M}_{\nu}
\right],
\ea
The normalization factor is
\ba{norm}
C_{0\nu} = (G_F^2 2 m_e)/(8\sqrt{2}\pi R).
\ea
Here, $m_e$ and $f_A \approx 1.261$ are the electron mass and
the nucleon axial coupling, $R = r_0 A^{1/3}$ is
the nuclear radius ($r_0 = 1.1 fm$).

The last term is the conventional Majorana neutrino mass contribution
proportional to the average neutrino mass. In the \rp MSSM we have
\ba{nu_ave}
\langle m_{\nu}\rangle = \sum_i m_{\nu_i} \left(V^{(\nu)}_{1i}\right)^2 =
m_{\nu_{\tau}}\left(V^{(\nu)}_{13}\right)^2  =
 \frac{2}{3}\frac{g_1^2 M_2}{Det M_{\chi}} \Lambda_1^2.
\ea
Here we neglected the small loop induced neutrino masses
$m_{\nu_e}\approx m_{\nu_{\mu}}\approx 0$ and used Eq. \rf{nu-mix}.

Let us specify the nuclear matrix elements involved in the formula for
the $\znbb$-decay amplitude \rf{R_tot}. They are
\ba{msq} \nn
{\cal M}_{\tilde g} &=& \left(\frac{m_{_A}}{m_p}\right)^2\frac{m_{_p}}{ m_e}
\left(M_{GT, {\tilde g}} + M_{T, {\tilde g}} \right),\\ \nn
%%%%%%%%%%%%%%%%%%%%%%%%%%%%%%%%%%%%%%%%%%%%%%%%%%%%%%%%%%%%%%%
~~~~~{\cal M}_{\nu} &=&
\left(\frac{f_V}{f_A}\right)^2 {\cal M}_{F,\nu} -
   {\cal M}_{GT,\nu}, \\
%%%%%%%%%%%%%%%%%%%%%%%%%%%%%%%%%%%%%%%%%%%%%%%%%%%%%%%%%%%%%%%
{\cal M}_{\lambda} &=& \frac{\alpha_P}{m_e R}
\left[ ~\frac{1}{3}{\cal M}_{GT,\lambda} +
{\cal M}_{T,\lambda}\right], \\ \nn
%%%%%%%%%%%%%%%%%%%%%%%%%%%%%%%%%%%%%%%%%%%%%%%%%%%%%%%%
{\cal M}_N &=&
\left(\frac{m_{_A}}{m_p}\right)^2\frac{m_p}{ m_e}\Big\{ \left(\frac{f_V}{f_A}\right)^2 {\cal M}_{F,N} -
   {\cal M}_{GT,N} \Big\}.
\ea
Here $m_p$ and $m_e$ stand for the proton and electron masses,
$f_V\approx 1.0$ is the vector nucleon constants,
$m_A = 0.85$GeV to be defined below.
The coefficient $\alpha_P = 1.75$ is related to the nucleon matrix
element of the pseudoscalar current. Its numerical value
calculated in the quark bag  model we take from Ref. \cite{HKK96}.

The partial nuclear matrix elements in the closure approximation we
write down in the form
\ba{short} \nn
{\cal M}_{F,i} &=& \langle 0^+_f| \sum^{}_{a\neq b}
{\tau}^{+}_{a}{\tau}^{+}_{b}
~{\cal F}_{i}(r_{ab})
\left(\frac{R}{r_{ab}}\right)^{\delta_i}
                                        |0^+_i\rangle , \\
%%%%%%%%%%%%%%%%%%%%%%%%%%%%%%%%%%%%%%%%%%%%%%%%%%%%%%%%%%%%%%%%%%%%%%%
{\cal M}_{GT,i} &=& \langle 0^+_f| \sum^{}_{a\neq b}
{\tau}^{+}_{a}{\tau}^{+}_{b}
~{\cal G}_{i}(r_{ab})
\left(\frac{R}{r_{ab}}\right)^{\delta_i}
~\sigma_{ab}~                                   |0^+_i\rangle , \\ \nn
%%%%%%%%%%%%%%%%%%%%%%%%%%%%%%%%%%%%%%%%%%%%%%%%%%%%%%%%%%%%%%%%%%%%%%%
{\cal M}_{T,i} &=& \langle 0^+_f| \sum^{}_{a\neq b}
{\tau}^{+}_{a}{\tau}^{+}_{b}
~ {\cal T}_i(r_{ab})
                  \left(\frac{R}{r_{ab}}\right)^{\delta_i}
~S_{ab}~                                      |0^+_i\rangle,
%%%%%%%%%%%%%%%%%%%%%%%%%%%%%%%%%%%%%%%%%%%%%%%%%%%%%%%%%
\label{(part)}
\ea
where
$i = \tilde g, \lambda, N, \nu$. The exponent
takes the values $\delta_i=\{1, 0, 1, 0\}$.
We use the shorthand notations
\ba{accumul}
{\cal F}_i &=& \{0, 0, F_N(x_A), h_+(r_{ab})\}, \ \ \
{\cal T}_i = \{F_{2}(x_{\pi}), h_{T'}(r_{ab}),  0, 0 \}, \\ \nn
{\cal G}_i  &=& \{F_{1}(x_{\pi}), h_R(r_{ab}), F_N(x_A), h_+(r_{ab})\},
\ea
for the following form factor functions and neutrino potentials
\ba{ffm}  \nn
F_1(x) &=& \left[\alpha^{1\pi} + \alpha^{2\pi}(x - 2)\right] e^{-x},\
F_N(x)  = \frac{x}{48} (3 + 3 x + x^2) e^{-x}, \\ \nn
F_2(x) &=&  \left[\alpha^{1\pi} \frac{3 + 3 x + x^2}{x^2}
+ \alpha^{2\pi}(x + 1)\right] e^{-x},\\
h_{+}(r_{ab}) &=& \frac{2}{\pi} R \int_0^{\infty}
dq\cdot q {\Phi}^2({\bf q}^2) \frac{j_0(q r_{ab})}{q + \bar A}, \\ \nn
%%%%%%%%%%%%%%%%%%%%%%%%%%%%%%%%%%%%%%%%%%%%%%%%%%%%%%%%%
h_{R}(r_{ab}) &=& \frac{2}{\pi} \frac{R^2}{m_p} \int_0^{\infty}
dq\cdot q^3{\Phi}^2({\bf q}^2) \frac{j_0(q r_{ab})}{q+ \bar A}, \\ \nn
%%%%%%%%%%%%%%%%%%%%%%%%%%%%%%%%%%%%%%%%%%%%%%%%%%%%%%%%%
%
\label{(Tt)}
h_{T'}(r_{ab}) &=&- \frac{2}{\pi} \frac{R^2}{3m_p} \int_0^{\infty}
dq\cdot q^3 {\Phi}^2({\bf q}^2) \frac{j_2(q r_{ab})}{q+ \bar A}.
\ea
with  $q = |\bf q|$ being an absolute value of the 3-momentum
transferred between the decaying nucleons.
$\alpha^{1\pi}=-4.4\cdot10^{-2}$ and
$\alpha^{2\pi}=0.2$ are the pion
structure coefficients introduced and calculated in Ref. \cite{FKSS97}.
$\bar A\approx 10$MeV is the average excitation energy of
the intermediate nuclear state.
The spherical Bessel functions are defined in the standard way
\ba{bes}
j_0(x) = \frac{\sin x}{x}, \ \ \
j_1(x) = \frac{\sin x}{x^2} - \frac{\cos x}{x}, \ \ \
j_2(x) =  \frac{3}{x}j_1(x) - j_0(x).
\ea
The nucleon form factor $\Phi(q^2)$ in Eqs. \rf{ffm}
takes into account the finite nucleon size. In our numerical analysis
we employ the conventional dipole parametrization
\ba{dip}
\Phi({\bf q}^2) = \left(1 + \frac{{\bf q}^2}{m_A^2} \right)^{-2}
\ea
with $m_A = 0.85$GeV.
We also defined
%The following notations are used
in Eqs. \rf{part}-\rf{Tt}:
\ba{notat}
S_{ab} &=&  3 \sir \sjr - \si \cdot \sj, \ \ \
\sigma_{ab} = \si \cdot \sj \nn \\
{\bfr}_{ab} &=& ({\bfr}_a -  {\bfr}_b ),\ \ \
r_{ab} = |{\bfr}_{ab}|, \ \ \ {\hat{\bfr}_{ab}^{~}}= {\bfr}_{ab}/r_{ab},\\
\nn
x_A &=& m_A r_{ab}, \ \ \ x_{\pi} = m_{\pi} r_{ab},
\ea
where ${\bf r}_a$ is the coordinate of the "$a$th" nucleon.

The following comments on the nuclear matrix element
${\cal M}_{\tilde g}$ in Eq. \rf{msq} associated with
the gluino graph Fig.1(b) are in order.
As discussed in Ref. \cite{FKSS97} it consists of the two parts
${\cal M}_{\tilde g} = {\cal M}_{\tilde g}^{2N} +
                       {\cal M}_{\tilde g}^{\pi N}$
corresponding to the gluino graph contribution via the two-nucleon
and the pion-exchange modes respectively. These two modes arise from
the two possibilities of hadronization of the 1st term of
the effective Lagrangian
${\cal L}_{eff}$ in Eq. \rf{eff}.
One can place the four quark fields present in this term
in the two initial neutrons and two final protons separately (2N-mode).
Then $nn\rightarrow pp + 2e^-$-transition is directly induced by
the underlying quark subprocess $dd\rightarrow uu + 2e^-$.
In this case the nucleon transition
is mediated by the exchange of a heavy particle which
is the gluino $\tilde g$
with the mass $m_{\tilde g}\geq 100$GeV. Therefore, the two decaying
neutrons are required to come up very closely to each other what
is suppressed by the nucleon repulsion.
Another possibility is to incorporate quarks involved in the underlying
\rp SUSY transition $dd\rightarrow uu + 2e^-$  not
into nucleons but into two virtual pions or into one pion as well as
into one initial neutron and one final proton \cite{FKSS97}.
Now  $nn\rightarrow pp + 2e^-$ transition is mediated by the charged
pion-exchange between the decaying nucleons ($\pi$N-mode).
Since the interaction region extends to the distances  $\sim 1/m_{\pi}$
this mode is not suppressed by the nucleon repulsion.
An additional enhancement of the $\pi$N-mode comes from
the hadronization of the \rp SUSY effective vertex operator
$\bar{u}\gamma_5 d\cdot\bar{u}\gamma_5 d\cdot \bar{e}P_R e^c$ replaced by
its hadronic image $\pi^2\cdot  \bar{e}P_R e^c$.
The enhancement occurs due to the coincidence
of the pseudoscalar quark bilinears $\bar{u}\gamma_5 d$ with
$\pi$-meson field.
As is shown in Ref. \cite{FKSS97} the $\pi$N-mode absolutely dominates over
the 2N-mode.  Therefore we neglected the subdominant 2N-mode part
${\cal M}_{\tilde g}^{2N}$ in Eq. \rf{msq}.

%%%%%%%%%%%%%%%%%%%%%%%%%%%%%%%%%%%%%%%%%%%%%%%%
%
We calculate the nuclear matrix elements within the renormalized Quasiparticle
Random Phase Approximation (pn-RQRPA) \cite{pn-RQRPA}. This nuclear
structure method has been developed from the proton-neutron QRPA
approach, which has been frequently used in the $\znbb$-decay calculations.
The pn-RQRPA is an extension of the pn-QRPA by incorporating the Pauli
exclusion principle for the fermion pairs.

The limitation of the conventional pn-QRPA is traced to the quasiboson
approximation (QBA), which violates the Pauli exclusion principle.
In the QBA one neglects the terms coming from the commutator of the
two bifermion operators by replacing the exact expression for this commutator
with its expectation value in the uncorrelated BCS ground state.
In this way the QBA implies the two-quasiparticle operator to be
a boson operator. The QBA leads to too strong ground state correlations
with increasing strength of the residual interaction in the
particle-particle channel what affects the calculated nuclear
matrix elements severely.

To overcome this problem the Pauli exclusion principle has to be
incorporated into the formalism \cite{pn-RQRPA} in order to
limit the number of quasiparticle pairs in the correlated ground state.
The commutator is not anymore boson like, but obtains corrections to
its bosonic behavior due to the fermionic constituents.
The pn-RQRPA goes beyond the QBA. The Pauli effect of
fermion pairs is included in the pn-RQRPA via the renormalized QBA (RQBA)
\cite{pn-RQRPA}, i.e.  by calculating the commutator of two bifermion
operators in the correlated RPA ground state.
Now it is widely recognized that the QBA  is a poor approximation and
that the pn-RQRPA offers the advantages over pn-QRPA. Let us stress that
there is no collapse of the pn-RQRPA solution for a physical value of
the nuclear force and that the nuclear matrix
elements have been found significantly less sensitive to the increasing
strength of particle-particle interaction in comparison with QRPA results.
Thus, the pn-RQRPA provides significantly more reliable treatment of
the nuclear many-body problem for the description of the
$0\nu\beta\beta$ decay.

For numerical treatment   of the $0\nu\beta\beta$-decay
matrix elements listed in Eqs. \rf{part} within the pn-RQRPA we transform
them by using the second quantization formalism
to the form containing the  two-body matrix elements
in the relative coordinate. One obtains \cite{si97}:
\begin{eqnarray}
<O_{12}> =
\sum_{{p n p' n' } \atop {J^{\pi}
m_i m_f {\cal J}  }}
~(-)^{j_{n}+j_{p'}+J+{\cal J}}(2{\cal J}+1)
\left\{
\matrix{
j_p &j_n &J \cr
j_{n'}&j_{p'}&{\cal J}}
\right\}\times~~~~~~~~\nonumber \\
<p, p';{\cal J}|f(r_{12})\tau_1^+ \tau_2^+ {\cal O}_{12}
f(r_{12})|n ,n';{\cal J}>\times ~~~~~~~~~~~~~
\nonumber \\
< 0_f^+ \parallel
\widetilde{[c^+_{p'}{\tilde{c}}_{n'}]_J} \parallel J^\pi m_f>
<J^\pi m_f|J^\pi m_i>
<J^\pi m_i \parallel [c^+_{p}{\tilde{c}}_{n}]_J \parallel
0^+_i >.
\end{eqnarray}
${\cal O}_{12}$ represents the coordinate and spin dependent part of the
two body transition operator of the $0\nu\beta\beta$-decay
nuclear matrix elements in Eqs. \rf{part}.
The short-range correlations
between the two interacting nucleons are taken into account by
a correlation function
\begin{equation}
\label{2nuccorr}
f(r)=1-e^{-\alpha r^2 }(1-b r^2) \quad \mbox{with} \quad
\alpha=1.1~ \mbox{fm}^2 \quad \mbox{and} \quad  b=0.68 ~\mbox{fm}^2.
\end{equation}
The one-body transition densities and other details of the
nuclear structure model are given in \cite{pn-RQRPA,si97}.

\begin{table}[t]
\caption{Nuclear matrix elements for
the neutrinoless double beta decay
 $^{76}Ge(0^{+}) \rightarrow ^{76}Se({0^{+}})$
within the pn-RQRPA.}
\label{table1}
\begin{tabular}{|cccccc|}
\hline
${\cal M}_{GT,N}$&${\cal M}_{F,N}\ \ $&${\cal M}_{GT,\nu}$&
${\cal M}_{F,\nu}\ \ $
&${\cal M}_{GT,\lambda}$ & ${\cal M}_{T,\lambda}\ \ $ \\
\hline
&&&&& \\
$0.071$ & $-0.025 $  & $2.6 \ \ \ $& $ -1.2 \ \ \ $ &
$1.20$&$0.21$\\
&&&&& \\
\hline
\hline
${\cal M}_{GT,{\tilde g}}$& ${\cal M}_{T, {\tilde g}}$&
${\cal M}_{\tilde g}$ & ${\cal M}_{\lambda}$ & ${\cal M}_{\nu}$   &
${\cal M}_{N}$\\
\hline
&&&&& \\
$ -0.34$ &$-0.089$ &$-649$ & $ 88 $ &$-3.4 $ & $ -132 $\\
& & & &&\\
& & & &&\\
\hline
\end{tabular}
\end{table}
The calculated nuclear matrix elements for the $0\nu\beta\beta$-decay
of A=76 isotope within the pn-RQRPA are presented
in Table 1. The considered single-particle
model space has been  the 12-level model space
(the full $2-4\hbar\omega$ major oscillator shells)
introduced in Ref.\cite{si97}.
The  nuclear matrix elements listed
in the Table 1 have been obtained for the $g_{pp} = 1.0$
where $g_{pp}$ is introduced to renormalize the particle-particle
interaction strength of the nuclear Hamiltonian.

According to our numerical analysis,  variations of
the nuclear matrix elements
${\cal M}_{\tilde g},{\cal M}_{\lambda}, {\cal M}_{N}$ and
${\cal M}_{\nu}$  do not exceed 15\% and 30\% respectively
within the physical region of the nuclear structure parameters.

Having all the quantities in the $\znbb$-decay amplitude Eq. \rf{R_tot}
specified we are ready to extract the limits on the \rp parameters
from the non-observation of the $\znbb$-decay.

\section{$\znbb$-decay constraints on bilinear R-parity violation.}
Starting from the Eq. \rf{R_tot} we derive the half-life formula
\ba{half-life}
\big[ T_{1/2}^{\znbb}(0^+ \rightarrow 0^+) \big]^{-1}
= G_{01} |{\cal M}_{\nu}|^2 |{\cal A}|^2.
\ea
Here $G_{01}$ is the phase space factor tabulated for various isotopes
in Ref. \cite{pa96}.

We introduced the dimensionless parameter
\ba{Aa}
{\cal A} =
\frac{\lg m_{\nu}\rg}{m_e} +
\frac{m_p}{<M_{\chi}>}\ \omega_{N} +
\eta_{\tilde g}\ \omega_{\tilde g} +
\lambda_{111}^{\p}\ \eta_{\nu}\ \omega_{\lambda},
\ea
where $ \omega_i = {\cal M}_i/{\cal M}_{\nu}$  with
$i = \tilde g, \lambda, N$.
The first, second and third terms in  this equation
correspond to the contributions of the neutrino, neutralino
and gluino graphs in Fig. 1(a,b).
Graphs in Fig. 1(c,d) contribute to the last term in Eq. \rf{Aa}.
It is worthwhile noticing that at typical randomly sampled values of
the MSSM parameters $M_2, \mu, \tbt$ the neutrino exchange
contribution from Fig. 1(a) dominates over the other contributions.

The most stringent experimental lower limit on the $0\nu\beta\beta$-decay
half-life has been obtained for $^{76}$Ge \cite{hdmo97}
\ba{exp}
T_{1/2}^{{0\nu\beta\beta}-\mbox{exp}}(0^+ \rightarrow 0^+)
\hskip2mm \geq \hskip2mm
1.1 \times 10^{25} \mbox{ years} \ \ \ \ \ \ \ \ \ 90 \% \ \mbox{c.l.}
\ea

With the  nuclear matrix elements calculated in the previous section
this lower limit can be cast into the following upper bound
\ba{A-lim}
|{\cal A}| \leq 1.0\cdot 10^{-6}.
\ea
This constraint represents a complex exclusion condition
placed by the non-observation of the $\znbb$-decay on
the \rp MSSM parameter space.
The individual bounds on the bilinear \rp   parameters $\mu_1, \lg\nu_1\rg$
of our present concern depend on concrete SUSY model
settings which fix the values of $M_2, \mu$ and $\tbt$ in
the left hand side of Eq. \rf{A-lim}.

Typical constrains for the 1st generation \rp parameters
$\mu_1, \lg\nu_1\rg, \mu_1\lambda^{\p}_{111},
                     \lg\nu_1\rg\lambda^{\p}_{111}$
can be obtained at the typical weak scale values of the MSSM parameters
$
M_2 = \mu = 100GeV
$
and $\tbt = 1$.
We also assume, as is commonly done in the similar cases, the absence of
a significant cancellation between the terms in the left hand side
of Eq. \rf{A-lim} defined in Eq. \rf{Aa}.
Thus, we come up with the following constraints
\ba{const_num}
|\mu_1| \leq 470 \mbox{KeV}, \\
\label{(vev)}
|\lg\tilde\nu_1\rg| \leq 840 \mbox{KeV},\\
|\mu_1\lambda^{\p}_{111}| \leq 100 \mbox{eV},\\
\label{(vev1)}
|\lg\tilde\nu_1\rg\lambda^{\p}_{111}| \leq 55 \mbox{eV}
\ea
Recall that in our notations $\lg\tilde\nu_1\rg \equiv \lg\tilde\nu_e\rg$.
To our knowledge these stringent constraints for the 1st generation
\rp   parameters were not previously considered in the literature
except a parenthetic note in Ref. \cite{hall}.
One can find in the published papers
%in the literature
only those constraints which
involve the combinations of the 1st and 2nd generation bilinear
\rp parameters \cite{Now} or contain only the 3rd generation ones
\cite{Mukh}.

To see how stringent are the obtained constraints
we can compare them with the following one
\ba{l_111}
\lambda^{\p}_{111} &\leq & 1.3\cdot 10^{-4}
\ea
which is known as a most stringent constraint on the R-parity
violation \cite{FKSS97}.
This constraint was previously obtained from the $\znbb$-decay
by taking into account only the superpotential trilinear couplings
in Eq. \rf{lambda}.
Consider for comparison the dimensionless quantity
\ba{scale}
\lambda^{LH}\approx  g_2^2 \ \frac{\mu_1\ or \  \lg\tilde\nu_1\rg}{M_{SUSY}},
\ea
with $M_{SUSY}\sim 100$GeV being the typical SUSY breaking scale.
As follows from Eqs. \rf{LH-vert}-\rf{coeff}
this dimensionless quantity sets the strength of
the RPM induced trilinear fermion-sfermion-fermion interaction
similarly to the coupling $\lambda_{111}$ in Eq. \rf{lambda}.
This makes reasonable the comparison of the constraints placed on
these couplings by the experiment.
From Eqs. \rf{const_num}-\rf{vev} we get an estimation
\ba{eff_const}
\lambda^{LH} \leq 10^{-6}.
\ea
This constraint looks more stringent (if such a comparison is
legitimate) than that for $\lambda_{111}$ in Eq. \rf{l_111}.

After all we conclude that the R-parity violation within
the 1st generation is restricted by the $\znbb$-decay to
a very low level. Now this statement holds for the generic case
of the \rp SUSY including both the superpotential trilinear
couplings and the bilinear terms in the superpotential as well as
in the soft supersymmetry breaking sector.

This conclusion has some immediate phenomenological consequences for
the other experiments, in particular for the accelerator ones.
For instance, among the two body decay modes of the neutralinos
\ba{accel}
\chi\longrightarrow e^{\pm} W^{\mp}, \mu^{\pm} W^{\mp},
\tau^{\pm} W^{\mp}, \ \ \
\chi\longrightarrow \nu_{e,\mu,\tau} Z,
\ea
and similar processes open in the presence of the bilinear \rp terms
one can now safely neglect the modes with electron or $\nu_e$.

We can also generalize the arguments used in the \rp SUSY interpretation
of the HERA anomaly \cite{hera}. It is believed that this anomaly can be
explained by the s-channel squark exchange
$ q_1 e \rightarrow \tilde q_i^* \rightarrow
q_je, \chi q_i, \chi^{\pm} q_i'$ between the initial quark-lepton state
and the final state particles. The quark-lepton vertex
$q \tilde q e$ allowed in the \rp SUSY models receives the contributions
both from the trilinear $\lambda^{\p}$ couplings and from
the trilinear operators induced by the bilinear terms via
the lepton-gaugino-higgsino mixing.
It is a common practice to neglect the 1st generation
squarks in the above mentioned \rp SUSY explanation of the HERA anomaly.
The argument is derived from the stringent constraint
on the 1st generation $\lambda^{\p}_{111}$ coupling \cite{FKSS97}
shown in Eq. \rf{l_111}.
However, it does not take into account the effect of the bilinear \rp
operators.
Now, having at hand the new stringent limit on the 1st generation
bilinear R-parity violation in Eq. \rf{const_num}-\rf{vev1}  we can extend
the validity of this argument to a general case of R-parity
violation considered in the present paper.

\section{Conclusion.}

In summary,  we derived the contribution of the bilinear
R-parity violating terms to the neutrinoless double beta decay.
Alone with the analysis of the trilinear terms previously made in
Refs. \cite{Mohapatra}-\cite{FKSS97} this completes the derivation
of all possible tree-level  contributions to the $\znbb$-decay within the \rp MSSM.

From the non-observation of $\znbb$-decay we obtained new stringent
upper limits on the 1st generation R-parity violating parameters
such as the lepton-Higgs mixing mass parameter $\mu_1$ and
the vacuum expectation value of the electron sneutrino
$\lg\tilde\nu_e\rg$.  Then we discussed some implications of these
constraints on the other experiments and, in particular, on
those which are running or planned at accelerators. We conclude that
the R-parity violating effects within the 1st generation, if exist,
are very small and in most cases can be neglected in
phenomenological analysis of observable effects.

A special attention  was paid to the effects of the nuclear structure
in the $\znbb$-decay. In the framework of the pn-QRPA approach
we obtained the nuclear matrix elements which are stable with respect to
the variation of the nuclear model parameters within the physical domain.
Thus, we believe that our conclusions concerning the particle physics
side of the $\znbb$-decay do not suffer from
the nuclear structure uncertainties.

\vskip10mm
\centerline{\bf Acknowledgments}
\bigskip
We are grateful to V.A.~Bednyakov for helpful discussions.
S.K. would like to thank the "Deutsche Forschungsgemeinschaft"
for financial support.

\setcounter{section}{0}
\def\theequation{\Alph{section}.\arabic{equation}}
%\begin{appendix}
\setcounter{equation}{0}

\section{Appendix A}

Below we present the mass matrices of the neutral
and charged fermion sectors for the general case of
the bilinear R-parity violation within the MSSM
field contents.
\subsection{Neutral fermion mass matrix.}
In the two component Weyl basis
\ba{basis1}
\Psi_{(0)}^{\p T} &=& (\nu_i,\,\,-i\lambda^{\p },\,\,
-i\lambda _{3},\,\, \tilde{H}^0_1,\,\, \tilde{H}_2^0), \\
\ea
the mass term of the neutral fermions is
\ba{neut_L}
{\cal L}^{(0)}_{mass} = -{1 \over 2}\Psi_{(0)}^{\p T}{\cal M}_0
                           \Psi^{\p}_{(0)}\ +\
\mbox{H.c.}\, ,
\ea
The $7\times 7$ mass matrix has the distinct see-saw structure
\ba{neut_m}
{\cal M}_0=\left(\begin{array}{cc}
0 & m \\
m^T & M_{\chi} \end{array}\right).
\ea
with $3\times 4$ matrix
\ba{subm}
m=\left(\begin{array}{cccc}
-M_Z s_W c_{\beta} u_1    & M_Z c_W c_{\beta} u_1    & 0 & -\mu_1 \\
-M_Z s_W c_{\beta}  u_2   & M_Z c_W c_{\beta}  u_2  & 0 & -\mu_2 \\
-M_Z s_W c_{\beta}  u_3  & M_Z c_W c_{\beta}  u_3 & 0 & -\mu_3
\end{array}\right) .
\ea
originating from the \rp bilinear terms in the superpotential
and the soft SUSY breaking sector.

In Eq. \rf{neut_m} $M_{\chi}$ is the usual $4 \times 4$
the MSSM neutralino mass matrix in the basis
$\{-i\lambda^\prime, -i\lambda_3, \tilde{H_1}, \tilde{H_2} \}$
\ba{MSSM-chi}
   M_{\chi} = \left( \begin{array}{cccc}
                   M_1  &  0          & -M_Z s_W c_{\beta} &  M_Z s_W s_{\beta}  \\
                     0  &  M_2        &  M_Z c_W c_{\beta}  & -M_Z c_W s_{\beta}  \\
           -M_Z s_W c_{\beta}   &  M_Z c_W c_{\beta}  &  0          &  -\mu       \\
            M_Z s_W s_{\beta}   & -M_Z c_W s_{\beta}  & -\mu        &           0 \\
                    \end{array} \right).
\ea
Here $u_i = \langle\tilde\nu_i\rangle/\langle H_1\rangle$ and
$\tbt = \langle H_2\rangle/\langle H_1\rangle$ and
$s_W = \cw,\  c_W = \cw,\  s_{\beta} = \sbt,\  c_{\beta} = \cbt. $.

In the mass eigenstate basis defined as
\ba{mass_0}
\Psi_{(0)i} = \Xi_{ij} \Psi_{(0)j}^{\p},\ \ \ \ \ \ \
\ea
the $7\times 7$ neutral fermion mass matrix
${\cal M}_0$ in Eq. \rf{neut_m} becomes diagonal
\ba{diag}
\Xi^*{\cal M}_0\Xi^{\dagger} = Diag\{m_{\nu_i}, m_{\chi_k}\},
\ea
where $m_{\nu_i}$ and $m_{\chi_i}$ are the physical neutrino
and neutralino masses. For the considered case
of the tree level mass matrix
the only one neutrino has a non-zero mass $m_{\nu_1}=m_{\nu_2} =0,\
m_{\nu_3} \neq 0$.
\subsection{Charged fermion mass matrix.}
The mass term of the charged fermion sector has the following form
\ba{m_term}
{\cal L}^{(\pm)}_{mass} =
- \Psi_{(-)}^{\p T} {\cal M}_{\pm} \Psi_{(+)}^{\p} +\mbox{H.c.}
\ea
in the two component Weyl spinor basis
\ba{basis2}
\Psi_{(-)}^{\p T} &=& (e^-_L,\,\, \mu^-_L,\,\, \tau^-_L,\,\,
-i\lambda_{-},\,\, \tilde H^-_1),\\
\Psi_{(+)}^{\p T} &=& (e^+_L,\,\, \mu^+_L,\,\, \tau^+_L,\,\,
-i\lambda_{+},\,\, \tilde H^+_2).
\ea
The $5\times 5$ charged fermion mass matrix is
\ba{cf_matrix}
{\cal M}_{\pm}=\left(\begin{array}{cc}
M^{(l)} & E \\
E' & M_{\chi^{\pm}}
\end{array}\right),
\ea
where $M_{\chi^{\pm}}$ is the MSSM chargino mass matrix
\ba{chi_pm}
M_{\chi^{\pm}} = \left(\begin{array}{cc}
M & \sqrt{2} M_Z c_W s_{\beta}  \\
\sqrt{2}M_Z c_W c_{\beta}  & \mu
\end{array}\right) .
\ea
The sub-matrices $E$ and $E'$ lead to the chargino-lepton mixing.
They are defined as
\ba{sbm_E}
\displaystyle{E=\left(\begin{array}{cc}
\sqrt{2}M_Z c_W c_{\beta}  u_1 & \mu_1 \\
\sqrt{2}M_Z c_W c_{\beta}  u_2 & \mu_2 \\
\sqrt{2}M_Z c_W c_{\beta}  u_3 & \mu_3
\end{array}\right)},
\ea
and
\begin{equation} \label{m44}
E' = - \left(\begin{array}{ccc}
0 & 0 & 0 \\
M^{(l)}_{1i}u_i & M^{(l)}_{2i}u_i  & M^{(l)}_{3i}u_i
\end{array}\right),
\end{equation}
where $M^{(l)}$ is the charged lepton mass matrix.
In a good approximation it can be treated as a diagonal matrix
$M^{(l)} = Diag\{m^{(l)}_i\}$  with $m^{(l)}_i$ being
the physical lepton masses.
Also, one can safely neglect matrix $E'$ compared to the other
entries of the full mass matrix \rf{cf_matrix} taking into
account smallness of the lepton masses.

Rotation to the mass eigenstate basis
\ba{mass+}
\Psi_{(\pm)i} = \Delta_{ij}^{\pm} \Psi_{(\pm)j}^{\p},
\ea
casts the mass matrix in Eq. \rf{cf_matrix} to a diagonal form
\ba{diag+}
\left(\Delta^{-}\right)^*{\cal M}_{\pm}\left(\Delta^{+}\right)^{\dagger} =
Diag\{m^{(l)}_i, m_{\chi^{\pm}_k}\},
\ea
where $m^{(l)}_i$ and  $m_{\chi^{\pm}_k}$ are the physical charged lepton
and chargino masses.

\section{Appendix B}
Here we give a short account on the results of the approximate
diagonalization method used in our analysis.
\setcounter{equation}{0}
\subsection{Neutral fermion mixing matrix.}
To leading order in the small expansion parameters $\xi$ defined below,
an approximate form of the neutral fermion $7\times 7$
mixing matrix introduced in Eqs. \rf{mass}, \rf{mass_0} is \cite{Now}
\ba{neutral-mix}
\Xi^*=
\left(\begin{array}{cc}
V^{(\nu)T}(1 -{1 \over 2}\xi \xi^{\dagger}) & -V^{(\nu)T} \xi \\
N^*\xi^{\dagger} &  N^*(1 -{1 \over 2}\xi^\dagger \xi)
\end{array}\right),
\ea
Here
\ba{xi_0}
\xi_{i1} &=& \frac{g_1 M_2 \mu}{2\ Det M_{\chi}} \Lambda_i, \ \
\xi_{i2} = - \frac{g_2 M_1 \mu}{2\ Det M_{\chi}} \Lambda_i, \\
\xi_{i3} &=& \frac{\mu_i}{\mu} +
\frac{g_2(M_1 + \tan^2\theta_W M_2)\sbt\cw M_Z}{2\ Det M_{\chi}}
\Lambda_i, \\
\xi_{i4} &=& - \frac{g_2(M_1 + \tan^2\theta_W M_2)
\cbt\cw M_Z}{2\ Det M_{\chi}}
\Lambda_i,
\ea
with $i = 1,2,3$.
The determinant of the MSSM neutralino mass matrix \rf{MSSM-chi} is
\ba{detchi}
Det M_{\chi} = \s2bt M_W^2 \mu (M_1 + \tan^2\theta_W  M_2) -  M_1 M_2 \mu^2.
\ea

The $4\times 4$ matrix $N$ rotates the MSSM neutralino mass matrix
$M_{\chi}$ to the diagonal form
\ba{chi-diag}
N^* M_{\chi} N^{\dagger}= Diag\{m_{\tilde{\chi}_i}\}\,
\ea
where $m_{\tilde{\chi}_i}$ are the physical neutralino masses.
Thus, to leading order in $\xi$ the mixing within the neutralino sector
is described as in the MSSM by
\ba{neutralino-MSSM}
\chi_k = N_{kn} \chi^{\p}_n
\ea
with $\chi^{\p}_n = (-i\lambda^{\p}, -i\lambda_3,
\tilde{H}^0_1, \tilde{H}_2^0)$ being the weak basis.

The $3\times 3$ matrix $V^{(\nu)}$ rotates the RPM
induced effective neutrino mass matrix to the diagonal form
\ba{nu-rotat}
V^T_{\nu}\; m_{eff}\; V_{\nu} = Diag\{0, \; 0, \; m_{\nu}\} \, ,
\ea
The tree level expression for this mass matrix can
be found in Ref. \cite{Now}.
The only non-zero neutrino mass is given by
\ba{nu-tau-again}
m_{\nu}= g_2^2\frac{M_1 +  \tan^2\theta_W M_2}{4\ Det M_{\chi}}
\vert \vec{\Lambda} \vert^2\, .
\ea
where
\ba{vec-again}
\Lambda_i = \mu \lg\nu_i\rg - \lg H_1\rg \mu_i\ ,
\ea

Let us show an explicit form of the neutrino mixing matrix
\ba{nu-mix-exp}
V_{\nu}=\left(\begin{array}{ccc}
\cos \theta_{13} & 0 & -\sin \theta_{13} \\
\sin \theta_{23}\sin \theta_{13} & \cos \theta_{23} & \sin \theta_{23}
\cos \theta_{13} \\
\sin \theta_{13} & \sin \theta_{23} & \cos \theta_{13}\cos \theta_{23}
\end{array}\right) ,
\ea
where the mixing angles are expressed through the vector
$\vec{\Lambda}$ as follows:
\ba{tan-nu}
\tan \theta_{13} = -{\Lambda_1 \over
\sqrt{\Lambda_2^2 + \Lambda_3^2}}, \;\;\;\;\;
\tan \theta_{23} = {\Lambda_2 \over \Lambda_3} \, .
\ea
The mixing within the neutrino sector to leading order
in $\xi$ is described by
\ba{neutralin}
\nu_k = V^{(\nu)*}_{nk} \nu^{\p}_n
\ea
with $\nu^{\p}_n = (\nu_e, \nu_{\mu}, \nu_{\tau})$ being the weak basis.
\subsection{Charged fermion mixing matrix.}
To leading order in the small expansion parameters $\xi^L$ and $\xi^R$
defined below, an approximate form of the charged fermion $5\times 5$
mixing matrix introduced in Eqs. \rf{mass}, \rf{mass+} reads
\ba{Delta-}
\left(\Delta^{-}\right)^* =
\left(\begin{array}{cc}
V_L(1 -{1 \over 2} \xi^{L^*}\xi^{L^T}) & -V_L\xi^{L^*} \\
U^*\xi^{L^T} & U^*(1 -{1 \over 2} \xi^{L^T}\xi^{L^*})
\end{array}\right) ,
\ea
and
\ba{Delta+}
\left(\Delta^{+}\right)^{\dagger}=
\left(\begin{array}{cc}
(1 -{1 \over 2} \xi^{R^*}\xi^{R^T})V_R^{\dagger}  & \xi^{R^*}V^{\dagger} \\
-\xi^{R^T} V_R^{\dagger} & (1 -{1 \over 2} \xi^{R^T}\xi^{R^*})V^{\dagger}
\end{array}\right)
\ea
Here
\ba{xi-left}
\xi^{L^*}_{i1} = \frac{g_2}{\sqrt{2}\ Det M_{\chi^{\pm}}} \Lambda_i, \ \ \
\xi^{L^*}_{i2} = \frac{\mu_i}{\mu} -
\frac{g_2 \sbt\cw M_Z}{\mu\ Det M_{\chi^{\pm}}} \Lambda_i,
\ea
with $i = 1,2,3$ and
\ba{xi_r}
\xi^{R^*} = M^{(l)\dagger} \xi^{L^*} \left(M_{\chi^{\pm}}^{-1}\right)^T.
\ea
This matrix is much smaller than $\xi^L$ by the factor $m_l/M_{SUSY}$,
where $m_l$ and $M_{SUSY}$ are the lepton masses and the typical
SUSY breaking scale $M_{SUSY}\sim$100GeV.
Thus the mixing between $(e^+_L,\,\, \mu^+_L,\,\, \tau^+_L)$ and
$(-i\lambda_{+},\,\, \tilde H^+_2)$ described by the off diagonal
blocks of the $\Delta^+$ in Eq. \rf{Delta+} is small
and, therefore, neglected in our analysis.

In Eqs. \rf{Delta-}-\rf{Delta+} the determinant of
the MSSM chargino mass matrix is
\ba{det_chi+-}
Det M_{\chi^{\pm}} = M_2 \mu - \s2bt M_W^2.
\ea

The other matrices are defined as follows:
\ba{Def22}
&& U^* M_{\chi^{\pm}}V^{\dagger}= Diag\{m_{\chi^{\pm}_i}\},\\ \nn
&& V_L M^{(l)} V_R^{\dagger} = Diag\{m_{l_i}\},
\ea
with $M_{\chi^{\pm}}$ and $M^{(l)}$ are the MSSM chargino and charged leptons
mass matrices  defined in Appendix A while
$m_{\chi^{\pm}_i}$ and $m_{l_i}$ are the physical chargino and
the charged lepton masses.

%\newpage
%%%%%%%%%%%%%%%%%%%%%%% bibliomacros%%%%%%%%%%%%%%%%%%%%%%%%%%%%%%
\def\ijmp#1#2#3{{\it Int. Jour. Mod. Phys. }{\bf #1~}(19#2)~#3}
\def\pl#1#2#3{{\it Phys. Lett. }{\bf B #1~}(19#2)~#3}
\def\zp#1#2#3{{\it Z. Phys. }{\bf C #1~}(19#2)~#3}
\def\prl#1#2#3{{\it Phys. Rev. Lett. }{\bf #1~}(19#2)~#3}
\def\rmp#1#2#3{{\it Rev. Mod. Phys. }{\bf #1~}(19#2)~#3}
\def\prep#1#2#3{{\it Phys. Rep. }{\bf #1~}(19#2)~#3}
\def\pr#1#2#3{{\it Phys. Rev. }{\bf D #1~}(19#2)~#3}
\def\np#1#2#3{{\it Nucl. Phys. }{\bf B #1~}(19#2)~#3}
\def\mpl#1#2#3{{\it Mod. Phys. Lett. }{\bf #1~}(19#2)~#3}
\def\arnps#1#2#3{{\it Annu. Rev. Nucl. Part. Sci. }{\bf
#1~}(19#2)~#3}
\def\sjnp#1#2#3{{\it Sov. J. Nucl. Phys. }{\bf #1~}(19#2)~#3}
\def\jetp#1#2#3{{\it JETP Lett. }{\bf #1~}(19#2)~#3}
\def\rnc#1#2#3{{\it Riv. Nuovo Cim. }{\bf #1~}(19#2)~#3}
\def\ap#1#2#3{{\it Ann. Phys. }{\bf #1~}(19#2)~#3}
\def\ptp#1#2#3{{\it Prog. Theor. Phys. }{\bf #1~}(19#2)~#3}
%%%%%%%%%%%%%%%%%%%%%%%%%%%%%%%%%%%%%%%%%%%%%%%%%%%%%%%%%%%%%%%%%

\newpage

\bigskip
{\large\bf Figure Captions}

\bigskip
\noindent
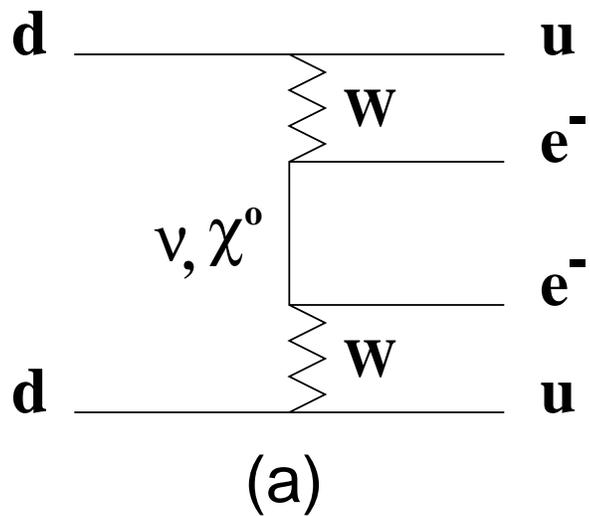
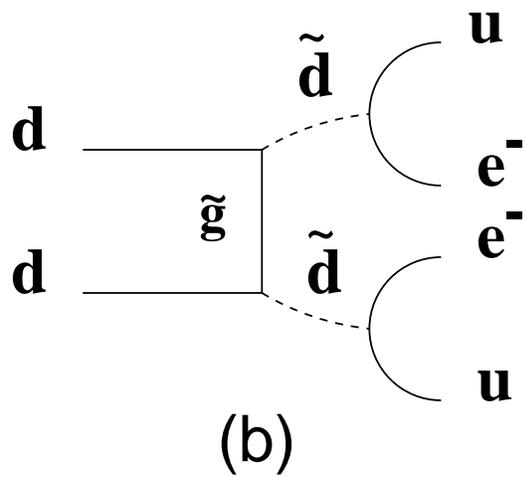
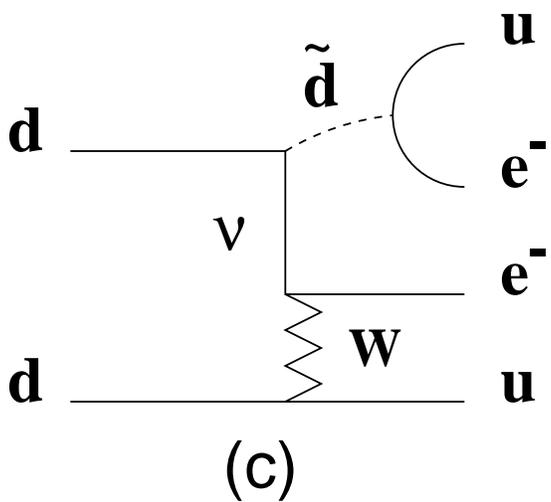
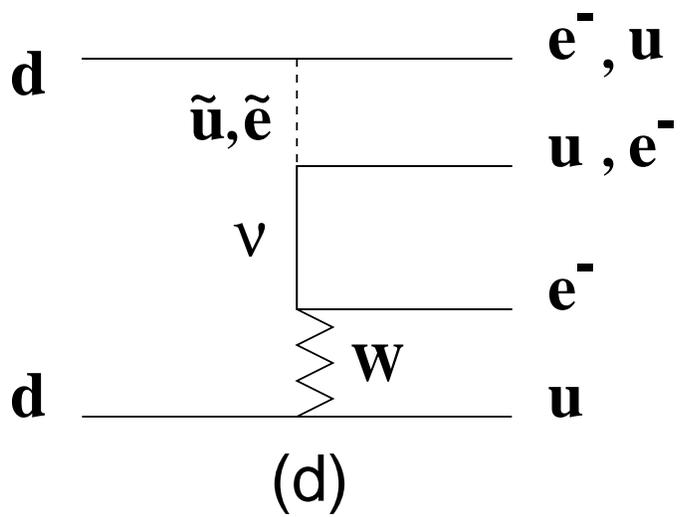
{\bf Fig.1.:}
Feynman graphs contributing to neutrinoless double beta ($\znbb$)
decay for the case of the bilinear R-parity violation.
(a) the Majorana neutrino or neutralino exchange with
two accompanying W-bosons; (b) the gluino-squark-squark
exchange;
(c,d) the neutrino-squark/slepton exchange with one accompanying W-boson.

\end{document}